\documentclass[a4paper, fleqn, usenatbib, useAMS]{mnras}

\def\aap{AA}
\def\apjl{ApJL}
\def\apjs{ApJS}
\def\mnras{MNRAS}
\def\apj{ApJ}

\def\nat{Nat}

\usepackage{afterpage}
\usepackage{multicol}        
\usepackage{graphicx}
\usepackage{float}
\usepackage{bm} 
\usepackage{amssymb}
\usepackage{hyperref}
\usepackage{amsmath} 
\usepackage{subfig} 
\usepackage{pdflscape}
\usepackage[export]{adjustbox}
\usepackage{mathrsfs} 
\usepackage{mathtools}

\title[Giant cold satellites from low-concentration haloes]{Giant cold satellites from low-concentration haloes}

\author[N. C. Amorisco]{Nicola C. Amorisco$^{1, 2}$\thanks{E-mail:
nicola.amorisco@cfa.harvard.edu}\\
$^{1}$Max Planck Institute for Astrophysics,  Karl-Schwarzschild-Strasse 1, 85748 Garching, Germany \\
$^{2}$Institute for Theory and Computation,  Harvard-Smithsonian centre for Astrophysics,  60 Garden St.,  Cambridge,  MA 02138,  USA}

\begin{document}



\maketitle

\label{firstpage}

\begin{abstract}
The dwarf satellite galaxies of the Milky Way Crater~II and Antlia~II have uncommonly low dynamical mass densities, 
due to their large size and low velocity dispersion. Previous work have failed to identify formation scenarios 
within the $\Lambda$CDM framework and have invoked cored dark matter haloes, processed by tides. 
I show that the tidal evolution of $\Lambda$CDM NFW haloes is richer than previously recognised: 
tidal heating causes the innermost regions of haloes that fall short of the mass-concentration relation to expand significantly, 
resulting in the formation of giant, kinematically cold satellites like Crater~II and Antlia~II.
Furthermore, while the satellite is reaching apocenter, extra-tidal material can cause an 
even more inflated appearance. When present, as likely for the larger Antlia~II, nominally unbound material can be 
recognised thanks to its somewhat hotter kinematics and line-of-sight velocity gradient. 
Contrary to other formation scenarios, Crater~II and Antlia~II may well have experienced very little mass loss,
as in fact hinted by their observed metallicity. 
If indeed a satellite of NGC1052, tidal evolution of a low-concentration halo may similarly have led to the formation of 
NGC1052-DF2.
\end{abstract}

\begin{keywords}
galaxies: dwarf --- galaxies: structure --- galaxies: evolution --- galaxies: haloes  ---  galaxies: kinematics and dynamics  \end{keywords}

\section{Introduction}

The satellite galaxies of the Local Group include the least luminous galaxies known as well as the most dark matter 
(DM) dominated \citep[e.g.][]{McC12,RE18,AF18}.
Their properties are indicative of the structure of low mass DM haloes and, as such, at the center of numerous studies 
aimed at deriving constraints on the nature of DM itself. 

The structural properties of two newly discovered `giant satellites', Crater II \citep[CraII, ][]{To16} and Antlia II \citep[AntII, ][]{To18}, 
have recently stirred the field. Both CratII and AntlII have a luminosity that is not much different from the one of 
Draco \citep[$L_{V,Draco}\approx 2.9\times10^5~L_\odot$, ][]{McC12} and close to the regime of the Classical dwarfs. 
However, both are extremely diffuse: with a half-light radius of $R_{\rm h}\sim1.1$~kpc, CratII has a surface brightness of $\mu\gtrsim30$; 
the half-light radius of AntlII is a surprising $R_{\rm h}\sim2.9$~kpc, for a surface brightness of $\mu\gtrsim32$. 
Both dwarfs are currently at a significant galactocentric distance and both are close to the apocenters of their orbits. 
Both are kinematically cold: the line of sight (LOS) velocity dispersion of AntII is of $\sigma=5.7\pm1.1$~kms$^{-1}$ \citep[][]{To18},
the one of CraII is of an impressing $\sigma=2.7\pm0.3$~kms$^{-1}$ \citep[][]{NC16}. AntII is a clear outlier in the 
size-luminosity plane \citep[][]{To18}, CraII a significant outlier in the size-velocity dispersion plane \citep[][]{NC16}.
The inferred dynamical mass density of both dwarfs within their half-light radius is accordingly and uncommonly low for their luminosity \citep[][]{NC16,To18}.

Recent works have found it impossible to reproduce such low dynamical mass densities within a $\Lambda$CDM framework \citep[][]{JS18,To18}.
On the one hand, it is well established that tidal evolution causes a progressive suppression of the velocity dispersion of the remnant 
\citep[][]{Ha03,JP08,JP10,JS18}, making a prolonged processing likely for AntII and CraII. On the other hand, this contrasts 
with their especially large size, as the same works have also shown that cosmologically motivated cuspy NFW haloes are 
not efficiently tidally heated and that the half-light radius of the satellites they host does not significantly grow during tidal evolution. 

This prompted the exploration of satellites hosted by haloes with cored density profiles, as if sculpted 
by baryonic feedback or caused by alternative DM models. These studies have revealed that cores are 
more efficiently heated by tides \citep[][]{RE15,JS18,To18} and 
a combination of a cored halo and significant tidal disruption has therefore been invoked to justify the properties of CraII and AntII.
In particular, \citet{JS18} and \citet{To18} propose that the progenitors of -- respectively -- CraII and AntII should have already had a 
significant size before infall and, most importantly, should have been significantly more luminous. 

In this Letter, I show that, like cored haloes, cuspy NFW haloes that fall short of the mass concentration relation 
are also more efficiently affected by tidal heating. Furthermore, the apparent size of these `giant satellites' can be
inflated by stellar material that is nominally unbound, but which remains spatially and kinematically coherent with 
the bound remnant itself. 
Section~2 collects a few preliminary considerations on tidal processing, Section~3 illustrates results from controlled 
numerical experiments, Section~4 discusses implications.

\section{Considerations on tidal heating}
The reference work on the tidal evolution of cuspy satellite galaxies around the Milky Way (MW) is \cite[][]{JP08}, in which
the authors use N-body experiments to study the evolution of a satellite NFW halo \citep[][]{NFW} 
orbiting in a potential that mimics the MW. 
The satellite has a virial mass of $M_{\rm vir}\sim4\times 10^9 M_\odot$ and an accordingly high concentration of $c=20$. 
Interestingly, the authors find that the time evolution of the structural properties of the bound remnant
depend only on the fraction of mass lost to tidal stripping. As this fraction increases, the velocity dispersion of the embedded 
satellite decrease while the effective radius evolves more slowly and eventually decreases.
This result has been broadly confirmed by a number of subsequent works, all of which have investigated the tidal evolution of low mass and, 
therefore, highly concentrated NFW haloes \citep[][]{JP10,RE15,JS18,To18}.
In turn, if featuring a central core, similar satellites are found to experience significant expansion of their innermost regions \citep[][]{RE15,JS18}.

I leave a detailed exploration of the mechanism underlying this dichotomy to future work, but briefly explain 
its origin here in terms of the phase space properties of a cuspy or cored halo, both approximately isotropic. 
In a cuspy NFW halo, i) the mean distance from the center $\langle r\rangle$ of individual particles goes to zero when proceeding 
deeper and deeper into the energy distribution of the system; ii) sub-systems composed of just the most bound $q\%$ of the 
same energy distribution are self-bound. The combination of these two facts makes the central regions of NFW haloes highly 
resilient to tidal processing and heating. In turn, the orbits of the most bound particles of a cored halo extend for a significant 
fraction of the core itself and tidal heating more efficiently affects the very bottom of the energy distribution of the system, 
causing even its innermost regions, where stars are hosted, to expand\footnote{The leading order of the kinetic energy 
gained by the individual particles of a system after the latter completes a pericentric passage is quadratic in the mean distance 
$\langle r\rangle$ \citep[e.g.,][]{JO72,LS87,ED10}.}.
Similarly, in NFW haloes with decreasing concentration, the energy injected by tidal shocks grows with the increasing $\langle r\rangle$,
while the binding energy of the most bound subset decreases. Cuspy NFW haloes with lower-than-average 
concentration should therefore be more susceptible to tidal heating, as I show in the following.

\section{Numerical Experiments}

I set up satellite haloes as isotropic spherical NFWs with a total of $N=4\times 10^6$ particles, and put them in orbit around a similar host
halo with $M_{\rm vir,host}=10^{12.2} M_\odot$ and $c_{\rm host}=12$. I use standard orbital properties for an average cosmological 
accretion event
around a MW mass halo at intermediate times \citep[e.g.,][]{Ji15}: with the notation of \citet{NA17}, I use $r_{\rm circ}/r_0=4$, where 
$r_{circ}$ is the radius of a circular orbit with the same energy, and $j=0.6$ where $j$ is the classical orbital circularity. 
I do not attempt to reproduce the orbital properties of AntII or CraII in detail, but this orbit remains broadly analogous to what 
inferred for these satellites and perhaps intermediate between the two, with an apocenter and pericenter of respectively 126 and 28~kpc.
All studied satellites have the same virial mass of $M_{\rm vir}=10^{9.9} M_\odot$, but I consider different values for the concentration: 
$c = (18, 9, 4.5)$.
The former is closer to the mean concentration for this mass at $z=0$, the others are nominally 
2.15$\sigma_{\log c}$ and 4.3$\sigma_{\log c}$ away from the mean, where $\sigma_{\log c}\approx~$0.14~dex \citep[][]{AL14}. 
Note, however, that the mean concentration of $\Lambda$CDM haloes decreases with increasing redshift \citep[$c=9$ 
would be closer to the mean at $z=1$, e.g.][]{AL14,AL16}. I let these haloes orbit for a total of 12 Gyr.

I consider five different values for the initial half-light radius $R_{\rm h,i}$ of the stellar component before infall, 
between $R_{\rm h,i}=180~$pc and $R_{\rm h,i}=800~$pc.
The simulations include no stellar particles and the dynamics of the stars is traced using those DM particles that are most bound before infall. 
Phase-space weights are a function of pre-infall energy alone and have a value of 1 for energies below some threshold $q$, or 0 above that
\citep[see e.g.][]{AC10}. Each pair of $(c, R_{\rm h,i})$ corresponds to a different tagging fraction $q$. 

\begin{figure*}
\centering
\includegraphics[width=.85\textwidth]{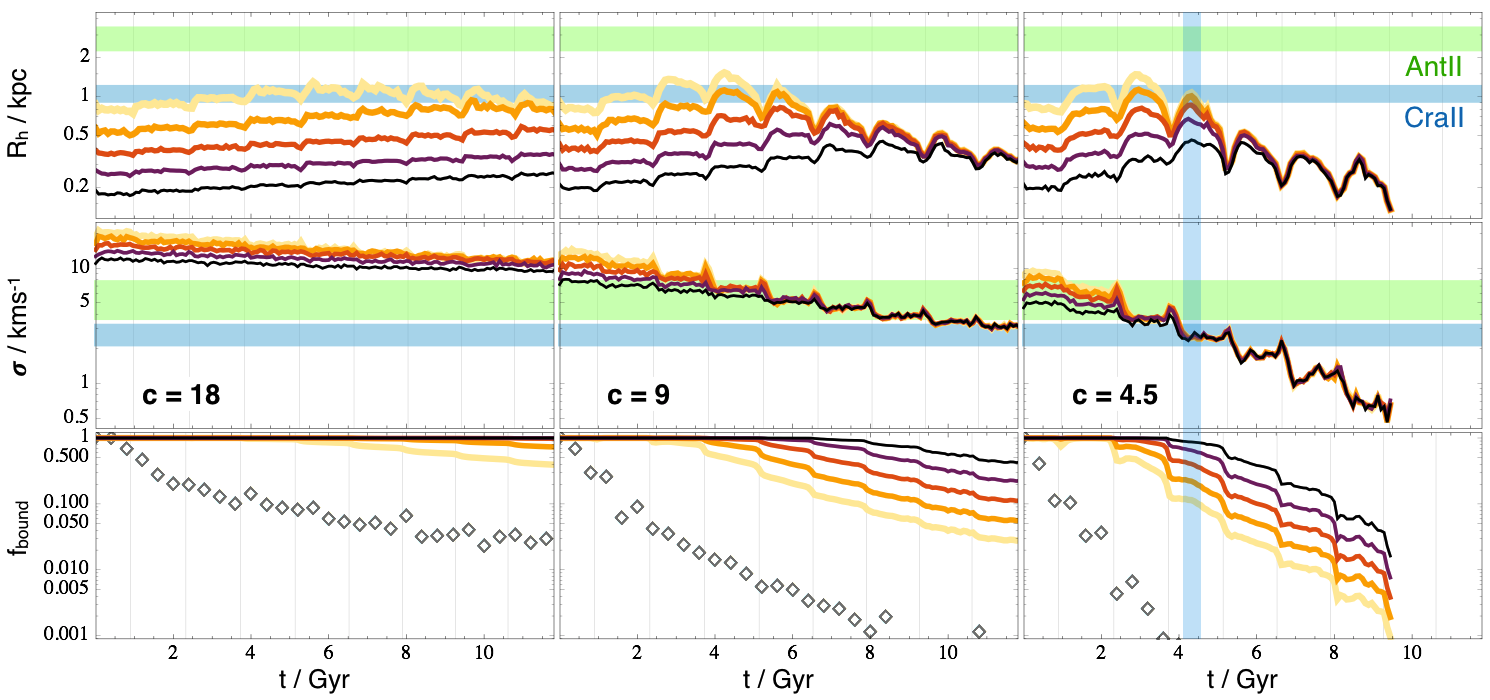}
\caption{The tidal evolution of the bound remnants in projected half light radius (upper panels), line-of-sight velocity dispersion (middle panels), bound fraction (lower panels), for haloes of different concentrations. The observer is located at the galactic centre. Lines of different colour refer to different values of the pre-infall half-light radius (see text for details).}
\end{figure*}

\subsection{Bound remnants}
Figure~1 displays the time evolution of the bound remnants. 
From left to right, columns refer to the three different concentrations. 
Upper panels show the half-light radius, $R_{\rm h}$. Middle panels the velocity dispersion, $\sigma$.
Lower panels show the bound fraction, in both stars (full lines) and DM (empty diamonds). In all panels, vertical grey lines identify pericentric passages,
lines of different colors refer to different values of $R_{\rm h,i}$. The shaded regions display 2$\sigma$ confidence
intervals for the observed properties of CraII (in cyan) and AntII (in green).
As time progresses, all three haloes continuously lose DM, as shown by the diamonds in the bottom panels, 
but the speed at which they do so is substantially different. The satellite hosted by the least concentrated halo is fully disrupted by the
end of the simulation, for any $R_{\rm h,i}$. If the satellite hosted by the $c=18$ halo has
$R_{\rm h,i}<400~$pc, it experiences no stripping by $t=12$~Gyr.
Both $R_{\rm h}$ and $\sigma$ assume an observer located at the galactic center, which is appropriate given the current location 
of CraII and AntII. In each column, satellites with larger $R_{\rm h,i}$ contain the particles tagged for lower $R_{\rm h,i}$, 
implying that values of $R_{\rm h}$ and $\sigma$ remain ordered at all times. 
As tides proceed to remove more and more bound particles, in an outside-in fashion, the five different curves converge 
to a unique time dependence. 

In agreement with previous work, the velocity dispersion is seen to progressively decrease. We find this is the case
independently of the initial concentration. Note however that the initial values of the velocity dispersion are substantially 
different due to the different concentrations. 
When $c=4.5$ and $c=9$, $R_{\rm h}$ is seen to efficiently expand after the first few pericentric passages. Then, as stellar mass 
begins to be lost, $R_{\rm h}$ decreases, `breathing' with orbital phase. The case $c=18$ appears 
to follow a qualitatively similar but much slower evolution. Equally high values of $R_{\rm h}$ are not achieved and, even when the system
is at its most expanded, $\sigma$ is too high to satisfactorily compare to AntII or CraII. 

As highlighted by the vertical shading in the right column, a bound remnant that can reproduce the properties of CraII is seen at $t\approx4.4$, 
after the third pericentric passage. At that time its velocity dispersion is approximately
independent of the pre-infall properties. In turn, in order to achieve a large enough $R_{\rm h}$, it must be $R_{\rm h,i}\gtrsim500~$pc, 
corresponding to a bound fraction $f_{\rm bound}\lesssim0.25$. No bound remnant as extended as AntII is seen in these simulations.

\begin{figure*}
\centering
\includegraphics[width=.9\textwidth]{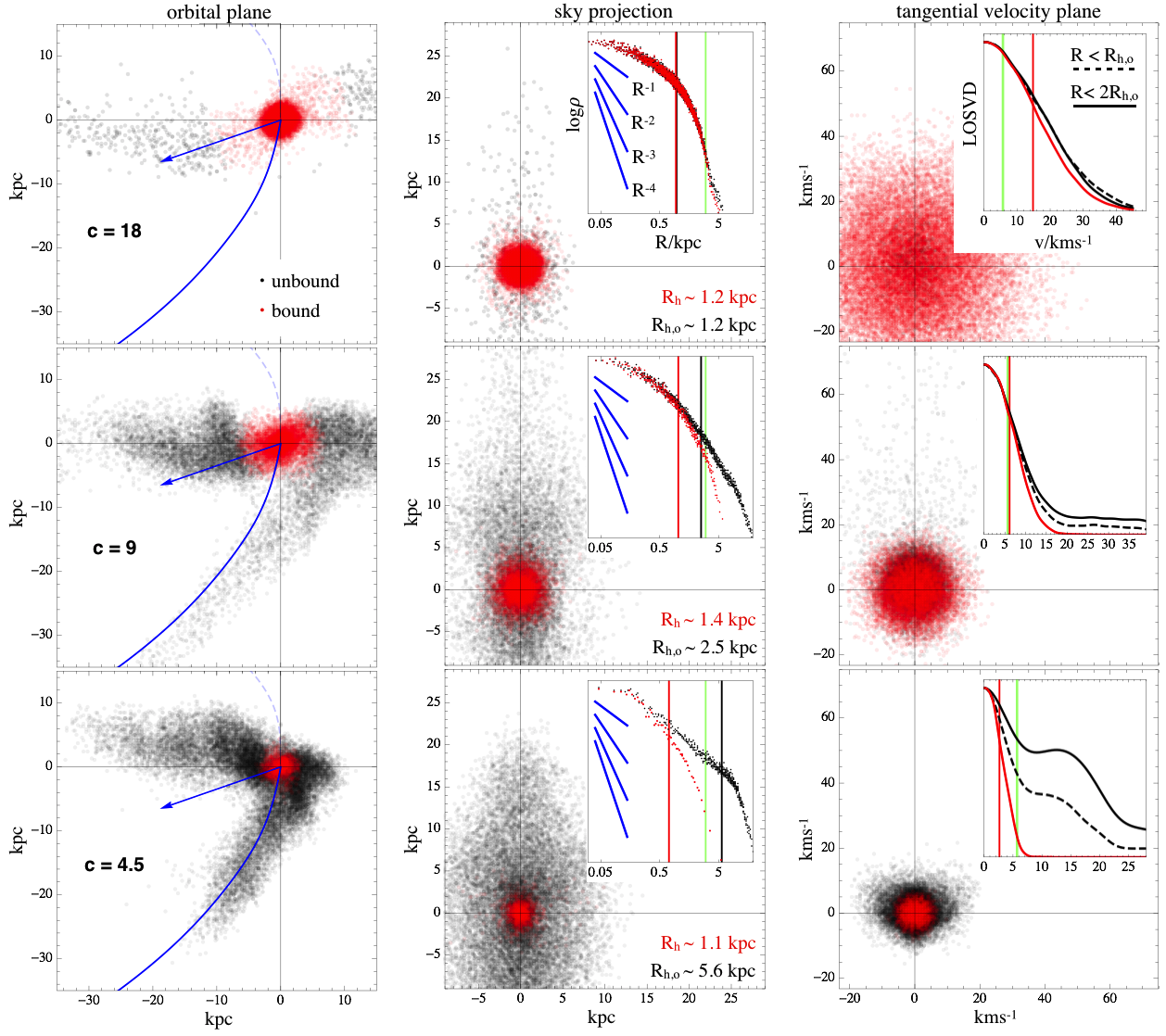}
\caption{Bound (red points) and unbound (black points) stellar material at $t=4.4$ Gyr, when the satellite is close to reaching apocenter after the third pericentric passage. Rows refer to haloes of different concentration, see the text for details. }
\end{figure*}

\subsection{Extratidal material but no extended stream}
Extra-tidal material is known surround the bound remnant in a diffuse halo that, in many cases, has a characteristic power-law 
dependence, $\rho\sim r^{-4}$ \citep[e.g.,][]{JB82,Ja87,JP10,JP17}.
As I will show, the dynamics of the extratidal material is different in the case of low-concentration haloes, especially when 
the satellite is about to reach apocenter. Unbound material remains spatially and kinematically coherent with the bound remnant,
further inflating its apparent size.

Figure~2 shows the three satellites at the time $t=4.4$ Gyr. Each row corresponds to a different value of the concentration. 
Left, middle and right columns show stellar particles as projected onto the satellite's orbital plane, the sky, and on the plane 
of tangential velocity. In the latter two columns orbital motion is along the vertical axis and the projection assume a viewer 
at the galactic center, which lies in the direction indicated by the blue arrow in the left panels.
The blue line in the same panels shows the satellite's orbit: the satellite is close to reaching apocenter. 
In all panels, stellar particles shown in red are nominally bound, black particles are unbound. 
For all three values of the concentration $R_{\rm h,i}=800~$pc.

The amount of unbound material grows with decreasing concentration, but what is more striking is that its location in the orbital plane depends
on the concentration too, despite the identical orbital phase. This is likely due the decreasing velocity (and energy) differences with respect 
to the remnant, which decrease the relative streaming speed when concentration decreases \citep[e.g.][]{KJ98,NA15}.  
The insets in the middle column show the projected density profile of the bound and all stellar particles 
(in red and black, respectively). As shown by the guiding blue lines, the unbound halo is as steep as expected when $c=18$, 
with a power-law slope between $-3$ and $-4$. Instead, the density profile of the extratidal material becomes more and more shallow 
in the less concentrated haloes, difficult to disentangle from the bound remnant.
The vertical black lines in the same insets show the `observed' half-light radius $R_{\rm h,o}$, whose value is reported at the bottom-right
of the middle panels: the unbound material significantly inflates the apparent size of the satellite, 
up to $R_{\rm h,o}=5.6~$kpc for the $c=4.5$ case!
As shown by the right column, the same unbound material remains highly coherent in tangential velocity, difficult to disentangle
from the remnant also based on proper motion. 
The insets in the right column show the line-of-sight velocity distribution (LOSVD), i.e. the distribution of radial velocities.
In all rows the red profile refers to the bound remnant (the red vertical line shows the corresponding velocity dispersion), 
the black dashed lines select all stellar material with $R<R_{\rm h,o}$, the full black line refers to the aperture $R<2R_{\rm h,o}$. 
Note that extratidal material becomes hotter in proceeding further out from the bound remnant, but part of this signal
may be lost when in the presence of other background contamination. 

Figure~3 focuses on the lowest concentration case, $c=4.5$, at the time $t=5.8$ Gyr, when the satellite is reaching
apocenter after its fourth pericentric passage. Differently from what done in Fig.~2, here I choose the initial properties of the 
stellar component so as to match the current $R_{\rm h}$ of CraII and AntII. This requires $R_{\rm h,i}\approx220~$pc
for CraII and $R_{\rm h,i}\approx350~$pc for AntII. As in Fig.~2, the upper-left panel shows density profiles, using red and black crosses
for the mock CraII system and red and grey points for the mock AntII system.
The lower-left panel shows the LOSVDs, in black for CraII and grey for AntII. Both are kinematically cold, featuring a bound core with 
$\sigma=1.7$kms$^{-1}$ and outskirts with gradually broader LOSVDs. Note however that the measured velocity dispersion 
will in fact depend on the adopted model and on the properties of any additional contamination. 
As shown by the upper-left panel, the mock AntII system includes a larger fraction of extratidal material which, 
as shown by the sky projection in the right panel, features a significant radial velocity gradient. This is a distinctive 
observational signature which can be sought with spectroscopic samples that extend extend well outside the observed 
half-light radius. 

\begin{figure}
\centering
\includegraphics[width=.85\columnwidth]{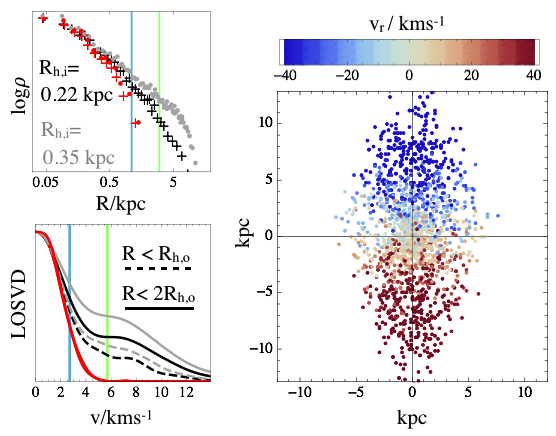}
\caption{Density profiles (upper-left panel) and line-of-sight velocity distribution (lower-left panel) of a mock CraII 
and AntII system, respectively in black and gray. The right panel shows a sky projection of the mock AntII system, illustrating 
the velocity gradient featured by the outer parts of the system.}
\end{figure}

Finally, note that in the models illustrated in Fig.~2 and~3 the `observed satellite' comprises 
essentially the entirety of the stellar material of the progenitor. In each case, a stream can be `added' by 
considering larger values of $R_{\rm h,i}$, but it is remarkable that, if formed from low-concentration haloes, 
systems like CraII and AntII may in fact have experienced very little mass loss and display no associated stream. 
This is different from what predicted in previous work, but in fact better matches 
the observed properties of both CraII and AntII, whose current stellar masses are approximately in line with what expected
based on their metallicity \citep{NC16,To18}.


\section{Discussion}

This Letter reinforces that tidally processed satellite galaxies may appear significantly different from what expected 
based on cosmologically motivated, isolated NFW haloes. Previous work has already shown that the central velocity dispersion 
of $\Lambda$CDM haloes on the mass-concentration relation decreases with time, which causes the dynamical mass density 
to decrease accordingly. This Letter shows that the spread in the achievable properties is wider than previously recognised, 
as haloes that fall short of the mass-concentration relation before infall also expand more efficiently. 
Given the significant width of the mass-concentration relation of $\Lambda$CDM haloes, this can be expected to have a 
substantial importance in shaping the properties of the population of current Local Group satellites \citep[see also][]{AF18}. 
In particular, tidal processing of low-concentration haloes naturally results in the formation of uncommonly large and kinematically 
cold bound remnants with especially low dynamical mass densities. When the satellite is close orbital apocenter the appearance of these remnants may be further inflated by unbound material. 
This provides a formation pathway for CraII and AntII \citep[but possibly also Andromeda XIX][and other Local Group satellites]{MC13,MC17} which remains fully within the $\Lambda$CDM framework, without requiring the presence of dark matter cores. 
\begin{figure}
\centering
\includegraphics[width=.95\columnwidth]{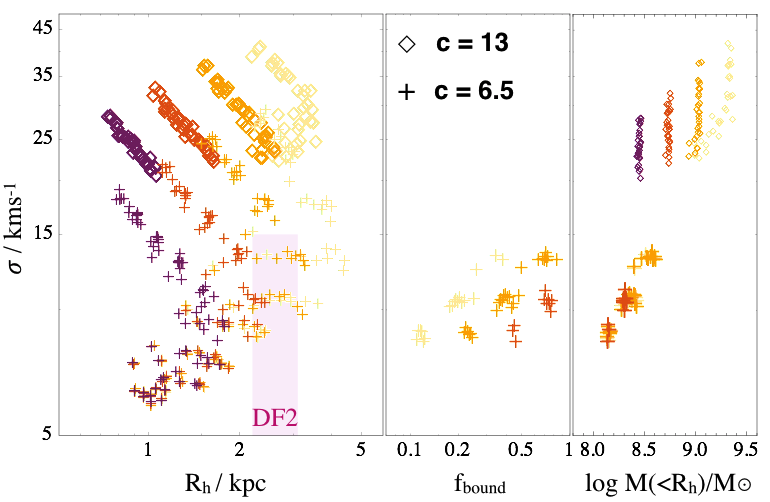}
\caption{Tidal evolution of the bound remnants in a mock NGC1052-DF2 system. }
\end{figure}

Due to their inclination to expand, the tidal processing of cored haloes has also been proposed as a possible formation 
scenario for satellite ultra-diffuse galaxies \citep[UDGs, ][]{TC18,GO18}, which, as AntII and CraII extend the galaxy 
population towards large sizes in the size-luminosity plane \citep[e.g.,][]{JK15,SD18}.
This Letter suggests that tidal evolution of low-concentration haloes may represent a useful line of investigation 
for UDG formation. In particular, as CraII and AntII, also NGC1052-DF2 has recently been at the centre 
of significant attention because of its possibly very low dynamical mass density \citep[][]{vD18}. This 
result is currently debated, with regard to both DF2's distance $D$ \citep[][]{vD18,IT18} or precise value of
the LOS velocity dispersion $\sigma$ \citep[][]{vD18,NM18,CL18,EM18,SD19}. Nonetheless, it is interesting
to note that the simulations presented here can broadly reproduce the possibly extreme properties of DF2, even if 
$D=20$~Mpc and $\sigma\lesssim15~$kpc. 

I rescale the host mass to $M_{\rm vir,host}=10^{13.2} M_\odot$ as approximately appropriate for NGC1052 \citep[e.g.,][]{GO18}. 
This brings the density profiles of two most concentrated satellites to match those of haloes with $M_{vir}=10^{10.9} M_\odot$ 
and concentrations $c=(13, 6.5)$. The left panel of Fig.~4 shows the evolution of their bound remnants in the plane 
$(R_{\rm h},\sigma)$, respectively with empty diamonds and crosses. In a number of snapshots the bound remnant
of the satellite hosted by the $c=6.5$ halo reproduces the assumed properties of DF2, shown by the shaded region. 
For these selected models the middle and right panels show the fraction of stars that is still bound to the remnant, $f_{\rm bound}$,
and the mass enclosed within $R_{\rm h}$, $M(<R_{\rm h})$. If the progenitor of DF2 was initially less concentrated than 
suggested by the mass-concentration relation, enclosed masses $8\lesssim \log M(<R_{\rm h})/M_\odot\lesssim8.5$ 
are not unfeasible. More closely tailored work will better address the case of DF2 as the present simulations do 
not include a self-gravitating stellar component. However, it remains useful to highlight here how DF2, CraII and AntII are 
likely connected, and how the tidal evolution of satellites hosted by low-concentration haloes may be responsible for 
the remarkable properties of all of these systems.

\section*{Acknowledgements}
It is a pleasure to thank Simon White and Adriano Agnello for inspiring discussions.


\begin{thebibliography}{99}

\bibitem[Amorisco(2015)]{NA15} Amorisco, N.~C.\ 2015, \mnras, 450, 575 
\bibitem[Amorisco(2017)]{NA17} Amorisco, N.~C.\ 2017, \mnras, 464, 2882 

\bibitem[Binney(1982)]{JB82} Binney, J.\ 1982, \mnras, 200, 951 


\bibitem[\protect\citeauthoryear{Caldwell, et al.}{2017}]{NC16} Caldwell N., et al., 2017, ApJ, 839, 20

\bibitem[Carleton et al.(2018)]{TC18} Carleton T., Errani R., Cooper M., Kaplinghat M., Pe{\~n}arrubia J., Guo Y., 2018, arXiv e-prints, arXiv:1805.06896

\bibitem[Collins et al.(2013)]{MC13} Collins, M.~L.~M., Chapman, S.~C., Rich, R.~M., et al.\ 2013, \apj, 768, 172 
\bibitem[Collins(2017)]{MC17} Collins, M.\ 2017, HST Proposal, 15302 

\bibitem[Cooper et al.(2010)]{AC10} Cooper, A.~P., Cole, S., Frenk, C.~S., et al.\ 2010, \mnras, 406, 744 



\bibitem[Danieli \& van Dokkum(2018)]{SD18} Danieli, S., \& van Dokkum, P.\ 2018, arXiv:1811.01962 

\bibitem[Danieli et al.(2019)]{SD19} Danieli, S., van Dokkum, P., Conroy, C., Abraham, R., \& Romanowsky, A.~J.\ 2019, arXiv:1901.03711 



\bibitem[D'Onghia et al.(2010)]{ED10} D'Onghia, E., Vogelsberger, M., Faucher-Giguere, C.-A., \& Hernquist, L.\ 2010, \apj, 725, 353 

\bibitem[Emsellem et al.(2018)]{EM18} Emsellem, E., van der Burg, R.~F.~J., Fensch, J., et al.\ 2018, arXiv:1812.07345 


\bibitem[Errani et al.(2015)]{RE15} Errani R., Penarrubia J., Tormen G., 2015, MNRAS, 449, L46
\bibitem[Errani et al.(2018)]{RE18} Errani R., Pe{\~n}arrubia J., Walker M.~G., 2018, MNRAS, 481, 5073

\bibitem[Fattahi et al. (2018)]{AF18} Fattahi, A., Navarro, J.~F., Frenk, C.~S., et al.\ 2018, \mnras, 476, 3816 


\bibitem[Hayashi et al.(2003)]{Ha03} Hayashi E., Navarro J.~F., Taylor J.~E., Stadel J., Quinn T., 2003, ApJ, 584, 541


\bibitem[Jaffe(1987)]{Ja87} Jaffe, W.\ 1987, Structure and Dynamics of Elliptical Galaxies, 127, 511 

\bibitem[Jiang et al.(2015)]{Ji15} Jiang, L., Cole, S., Sawala, T., \& Frenk, C.~S.\ 2015, \mnras, 448, 1674 

\bibitem[Johnston(1998)]{KJ98} Johnston, K.~V.\ 1998, \apj, 495, 297 




\bibitem[Koda et al.(2015)]{JK15} Koda, J., Yagi, M., Yamanoi, H., \& Komiyama, Y.\ 2015, \apjl, 807, L2 

\bibitem[Laporte et al.(2018)]{CL18} Laporte, C.~F.~P., Agnello, A., \& Navarro, J.~F.\ 2018, \mnras,  



\bibitem[Ludlow et al.(2014)]{AL14} Ludlow, A.~D., Navarro, J.~F., Angulo, R.~E., et al.\ 2014, \mnras, 441, 378 
\bibitem[Ludlow et al.(2016)]{AL16} Ludlow, A.~D., Bose, S., Angulo, R.~E., et al.\ 2016, \mnras, 460, 1214 

\bibitem[Martin et al.(2018)]{NM18} Martin, N.~F., Collins, M.~L.~M., Longeard, N., \& Tollerud, E.\ 2018, \apjl, 859, L5 



\bibitem[\protect\citeauthoryear{McConnachie}{2012}]{McC12} McConnachie A.~W., 2012, AJ, 144, 4

\bibitem[Navarro et al.(1996)]{NFW} Navarro, J.~F., Frenk, C.~S., \& White, S.~D.~M.\ 1996, \apj, 462, 563 

\bibitem[\protect\citeauthoryear{Ogiya}{2018}]{GO18} Ogiya G., 2018, MNRAS, 480, L106


\bibitem[Ostriker et al.(1972)]{JO72} Ostriker, J.~P., Spitzer, L., Jr., \& Chevalier, R.~A.\ 1972, \apjl, 176, L51 


\bibitem[Pe{\~n}arrubia et al.(2008)]{JP08} Pe{\~n}arrubia J., Navarro J.~F., McConnachie A.~W., 2008, ApJ, 673, 226
\bibitem[Pe{\~n}arrubia et al.(2010)]{JP10} Pe{\~n}arrubia J., Benson A.~J., Walker M.~G., Gilmore G., McConnachie A.~W., Mayer L., 2010, MNRAS, 406, 1290

\bibitem[Pe{\~n}arrubia et al.(2017)]{JP17} Pe{\~n}arrubia, J., Varri, A.~L., Breen, P.~G., Ferguson, A.~M.~N., \& S{\'a}nchez-Janssen, R.\ 2017, \mnras, 471, L31 


\bibitem[Sanders et al.(2018)]{JS18} Sanders J.~L., Evans N.~W., Dehnen W., 2018, MNRAS, 478, 3879


\bibitem[Spitzer(1987)]{LS87} Spitzer, L.\ 1987, Princeton, NJ, Princeton University Press, 1987, 191 p.,  

\bibitem[Torrealba et al.(2016)]{To16} Torrealba G., Koposov S.~E., Belokurov V., Irwin M., 2016, MNRAS, 459, 2370
\bibitem[Torrealba et al.(2018)]{To18} Torrealba G., et al., 2018, arXiv e-prints, arXiv:1811.04082

\bibitem[Trujillo et al.(2018)]{IT18} Trujillo, I., Beasley, M.~A., Borlaff, A., et al.\ 2018, arXiv:1806.10141 



\bibitem[van Dokkum et al.(2018)]{vD18} van Dokkum, P., Danieli, S., Cohen, Y., et al.\ 2018, \nat, 555, 629 

\end{thebibliography}
\end{document}